\documentclass[letterpaper]{article}
\usepackage{natbib,alifeconf}   
\usepackage{ctable, amsmath,amsfonts,amssymb,cite,graphicx,caption,amsthm,enumerate}
\pdfoutput=1

\newcommand{\cN}{{\mathcal N}}
\newcommand{\cP}{{\mathcal P}}

\newcommand{\vertices}{{V}}
\newcommand{\X}{{X}}
\newcommand{\Xin}{{X_{in}}}
\newcommand{\Xout}{{X_{out}}}
\newcommand{\xin}{{x_{in}}}
\newcommand{\xout}{{x_{out}}}
\newcommand{\occ}{{v}}
\newcommand{\mocc}{{\mathbf{U}}}
\newcommand{\grd}{{\mathbf{G}}}
\newcommand{\chn}{{\mathbf{C}}}
\newcommand{\Sit}{{S}}
\newcommand{\Act}{{A}}
\newcommand{\sit}{{s}}
\newcommand{\act}{{a}}

\newcommand{\mAct}{{\mathbf{A}}}

\newcommand{\coarse}{{\mathcal K}}

\newcommand{\fm}{{\mathfrak m}}

\newtheorem{step}{Step}
\newtheorem*{defn}{Definition}

\theoremstyle{remark}

\newtheorem*{rem}{Remark}

\title{Detecting emergent processes in cellular automata with excess information}
\author{David Balduzzi$^1$\\
\mbox{}\\
$^1$ Department of Empirical Inference, MPI for Intelligent Systems, T{\"u}bingen, Germany\\
david.balduzzi@tuebingen.mpg.de}

\begin{document}
\maketitle

\begin{abstract}
	Many natural processes occur over characteristic spatial and temporal scales. This paper presents tools for (i) flexibly and scalably coarse-graining cellular automata and (ii) identifying which coarse-grainings express an automaton's dynamics well, and which express its dynamics badly. We apply the tools to investigate a range of examples in Conway's Game of Life and Hopfield networks and demonstrate that they capture some basic intuitions about emergent processes. Finally, we formalize the notion that a process is emergent if it is better expressed at a coarser granularity.  
\end{abstract}

\section{Introduction}
\label{s:introduction}

Biological systems are studied across a range of spatiotemporal scales -- for example as collections of atoms, molecules, cells, and organisms \citep{Anderson:1972oq}. However, not all scales express a system's dynamics equally well. This paper proposes a principled method for identifying which spatiotemporal scale best expresses a cellular automaton's dynamics. We focus on Conway's Game of Life and Hopfield networks as test cases where collective behavior arises from simple local rules.

Conway's Game of Life is a well-studied artificial system with interesting behavior at multiple scales \citep{berlekamp:82}. It is a 2-dimensional grid whose cells are updated according to deterministic rules. Remarkably, a sufficiently large grid can implement any deterministic computation. Designing patterns that perform sophisticated computations requires working with distributed structures such as gliders and glider guns rather than individual cells \citep{dennett:91}. This suggests grid computations may be better expressed at coarser spatiotemporal scales.

The first contribution of this paper is a coarse-graining  procedure for expressing a cellular automaton's dynamics at different scales. We begin by considering cellular automata as collections of spacetime coordinates termed occasions (cell $n_i$ at time $t$). Coarse-graining groups occasions into structures called \emph{units}. For example a unit could be a $3\times3$ patch of grid containing a glider at time $t$. Units do not have to be adjacent to one another; they interact through \emph{channel} -- transparent occasions whose outputs are marginalized over. Finally, some occasions are set as \emph{ground}, which fixes the initial condition of the coarse-grained system. 

Gliders propagate at 1/4 diagonal squares per tic -- the grid's ``speed of light''. Units more than $4n$ cells apart cannot interact within $n$ tics, imposing constraints on which coarse-grainings can express glider dynamics. It is also intuitively clear that units should group occasions concentrated in space and time rather than scattered occasions that have nothing to do with each other. In fact, it turns out that most coarse-grainings  express a cellular automaton's dynamics badly. 

The second contribution of this paper is a method for distinguishing good coarse-grainings from bad based on the following principle:
\begin{itemize}
	\item
	\emph{Coarse-grainings that generate more information, relative to their sub-grainings, better express an automaton's dynamics than those generating less.}
\end{itemize}
We introduce two measures to quantify the information generated by coarse-grained systems. Effective information, $ei$, quantifies how selectively a system's output depends on its input. Effective information is high if few inputs cause the output, and low if many do. Excess information, $\xi$, measures the difference between the information generated by a system and its subsystems. 

With these tools in hand we investigate coarse-grainings of Game of Life grids and Hopfield networks and show that grainings with high $ei$ and $\xi$ capture our basic intuitions regarding  emergent processes. For example, excess information distinguishes boring (redundant) from interesting (synergistic) information-processing, exemplified by blank patches of grid and gliders respectively. 

Finally, the penultimate section converts our experience with examples in the Game of Life and Hopfield networks into a provisional formalization of the principle above. Roughly, we define a process as \emph{emergent} if it is better expressed at a coarser scale. 

The principle states that emergent processes are more than the sum of their parts -- in agreement with many other approaches to quantifying emergence \citep{crutchfield:94, tononi:04, polani:06, shalizi:06, Seth:2010ve}. Two points distinguishing our approach from prior work are worth emphasizing. First, coarse-graining is \emph{scalable}: coarse-graining a cellular automaton yields another cellular automaton. Prior works identify macro-variables such as temperature \citep{shalizi:06} or centre-of-mass \citep{Seth:2010ve} but do not show how to describe a system's dynamics purely in terms of these macro-variables. By contrast, an emergent coarse-graining is itself a cellular automaton, whose dynamics are computed via the mechanisms of its units and their connectivity (see below).

Second, our starting point is selectivity rather than predictability. Assessing predictability necessitates building a model and deciding what to predict. Although emergent variables may be robust against model changes \citep{Seth:2010ve}, it is unsatisfying for emergence to depend on properties of \emph{both} the process \emph{and} the model. By contrast, effective and excess information depend only on the process: the mechanisms, their connectivity, and their output. A process is then emergent if its internal dependencies are best expressed at coarse granularities.

\section{Probabilistic cellular automata}
\label{s:pca}

\subsubsection{Concrete examples.}
This paper considers two main examples of cellular automata: Conway's Game of Life and Hopfield networks \citep{hopfield:82}. 

The Game of Life is a grid of deterministic binary cells. A cell outputs 1 at time $t$ iff: (i) three of its neighbors outputted 1s at $t-1$ or (ii) it and two neighbors outputted 1s at $t-1$.

In a Hopfield network \citep{amit:89}, cell $n_k$ fires with probability proportional to 
\begin{equation}
	p(n_{k,t}=1|n_{\bullet,t-1})\propto \exp\left[\frac{1}{T}\sum_{j\rightarrow k}\alpha_{j k}\cdot n_{j,{t-1}}\right]
	\label{e:hopfield}
\end{equation}
Temperature $T$ controls network stochasticity. Attractors $\{\xi^1,\ldots,\xi^N\}$ are embedded into a network by setting the connectivity matrix as $\alpha_{j k}=\sum_{\mu=1}^N (2\xi_j^\mu-1)(2\xi_k^\mu-1)$.

\subsubsection{Abstract definition.}
A \emph{cellular automaton} is a finite directed graph $\X$ with vertices $\vertices_\X=\{\occ_1\ldots\occ_n\}$. Vertices are referred to as occasions; they correspond to \emph{spacetime} coordinates in concrete examples. Each occasion $\occ_l\in \vertices_X$ is equipped with finite output alphabet $\Act_l$ and Markov matrix (or \emph{mechanism}) $p_l(\act_l|\sit_l)$, where $\sit_l\in \Sit_l=\prod_{k\rightarrow l}\Act_k$, the combined alphabet of the occasions targeting $\occ_l$. The mechanism specifies the probability that occasion $\occ_l$ chooses output $\act_l$ given input $\sit_l$. The input alphabet of the entire automaton $\X$ is the product of the alphabets of its occasions $\Xin:=\prod_{l\in \vertices_\X} \Act_l$. The output alphabet is $\Xout=\Xin$. 

\begin{rem}
	The input $\Xin$ and output $\Xout$ alphabets are distinct copies of the same set. Inputs are causal interventions imposed via Pearl's $do(-)$ calculus \citep{pearl:00}. The probability of output $\act_l$ is computed via the Markov matrix: $p_l\big(\act_l|do(\sit_l)\big)$.
	The $do(-)$ is not included in the notation explicitly to save space. However, it is always implicit when applying any Markov matrix.
\end{rem}

A Hopfield network over time interval $[\alpha,\beta]$ is an abstract automaton. Occasions are spacetime coordinates -- e.g. $\occ_l=n_{i,t}$, cell $i$ at time $t$. An edge connects $\occ_k\rightarrow \occ_l$ if there is a connection from $\occ_k$'s cell to $\occ_l$'s and the time coordinates are $t-1$ and $t$ respectively for some $t$. The mechanism is given by Eq.~\eqref{e:hopfield}. Occasions at $t=\alpha$, with no incoming edges, can be set as fixed initial conditions or noise sources. Similar considerations apply to the Game of Life.

Non-Markovian automata (whose outputs depend on inputs over multiple time steps) have edges connecting occasions separated by more than one time step.

\section{Coarse-graining}
\label{s:coarse}

Define a \emph{subsystem} $\X$ of cellular automaton $Y$ as a subgraph containing a subset of $Y$'s vertices and a subset of the edges targeting those vertices. We show how to coarse-grain 
$\X$. 

\begin{defn}[coarse-graining]
	Let $\X$ be a subsystem of $Y$. The coarse-graining algorithm detailed below takes $\X\subset Y$ and data $\coarse$ as arguments, and produces new cellular automaton $\X_\coarse$. Data $\coarse$ consists of (i) a partition of $\X$'s occasions $\vertices_\X=\grd\cup \chn\cup \mocc_1\cup \cdots \cup \mocc_N$ into ground $\grd$, channel $\chn$ and units $\mocc_1\ldots\mocc_N$ and (ii) ground output $s^\grd$. 
\end{defn}

Vertices of automaton $\X_\coarse$, the new coarse-grained occasions, are units: $\vertices_{\X_\coarse} := \{\mocc_1\ldots \mocc_N\}$. The directed graph of $\X_\coarse$ is computed in Step 4 and the alphabets $\mAct_l$ of units $\mocc_l$ are computed in Step 5. Computing the Markov matrices (mechanisms) of the units takes all five steps.

The ground specifies occasions whose outputs are fixed: the initial condition $s^\grd$. The channel specifies unobserved occasions: interactions between units propagate across the channel. Units are macroscopic occasions whose interactions are expressed by the coarse-grained automaton. Fig.~\ref{f:coarse-graining} illustrates coarse-graining a simple automaton.

There are no restrictions on partitions. For example, although the ground is intended to provide the system's initial condition, it can contain any spacetime coordinates so that in pathological cases it may obstruct interactions between units. Distinguishing good coarse-grainings from bad is postponed to later sections.

\subsubsection{Algorithm.}
Apply the following steps to coarse-grain:

\begin{figure}[thpb]
	\centering
	\includegraphics[scale=0.8]{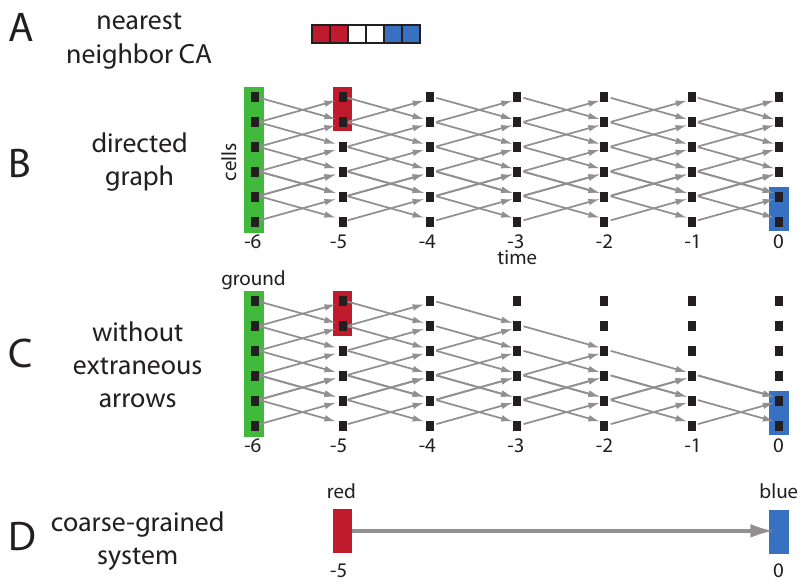}
	\caption{\footnotesize{
	(A) An automaton of 6 cells connected to their immediate neighbors. (B): The directed graph of occasions over time interval $[-6,0]$. Green occasions are ground. Red and blue occasions form two units. Other occasions are channel. (C): Edges  whose signals do not reach the blue unit have no effect. (D):  The coarse-grained system consists of two units (macro-occasions).   
	}}
	\label{f:coarse-graining}
\end{figure}

\begin{step}
	Marginalize over extrinsic inputs.
\end{step}

External inputs are treated as independent noise sources; we are only interested in \emph{internal}  information-processing. An occasion's input alphabet decomposes into a product $\Sit_l=\Sit_l^\X\times \Sit_l^{Y\setminus \X}$ of inputs from within and without the system. For each occasion $\occ_l\in \vertices_\X$, marginalize over external outputs using the uniform distribution:
\begin{equation}
	p_l\big(\act_l\big|\sit_l^\X) := 
	\sum_{\Sit_l^{Y\setminus \X}}p_l\big(\act_l\big|\sit_l^\X,\sit_l^{Y\setminus \X}\big)
	\cdot p_{unif}(\sit_l^{Y\setminus \X}).
	\label{e:noise}
\end{equation}

\begin{step}
	Fix the ground.
\end{step}

Ground outputs are fixed in the coarse-grained system. Graining $\coarse$  imposes a second decomposition onto $\occ_l$'s input alphabet, $\Sit_l^\X=\Sit_l^\grd\times \Sit_l^\chn\times \Sit_l^\mocc$ where $\mocc=\cup_k \mocc_k$. Subsume the ground into  $\occ_l$'s mechanism by specifying
\begin{equation*}
	p^\grd_l\big(\act_l\big|\sit_l^\chn, \sit_l^\mocc) := 
	p_l\big(\act_l\big|\sit_l^\grd, \sit_l^\chn, \sit_l^\mocc\big).
\end{equation*}

\begin{step}
	Marginalize over the channel.
\end{step}

The channel specifies transparent occasions. Perturbations introduced into units propagate through the channel until they reach other units where they are observed. Transparency is imposed by marginalizing over the channel occasions in the product mechanism
\begin{equation}
	p_\coarse(x_{out}^\coarse|x_{in}^\coarse):=\sum_{l\in \chn}
	\prod_{l\in \chn\cup \mocc}p^\grd_l\big(x_{out}^l|x_{in}^l\big),
	\label{e:u-mech}
\end{equation}
where superscripts denote that inputs and outputs are restricted, for $\coarse$, to occasions in units in $\coarse$ (since channel is summed over and ground is already fixed) and, for each $l$, to the inputs and outputs of occasion $\occ_l$.

For example, consider cellular automaton with graph $v_a\rightarrow v_b\rightarrow v_c$ and product mechanism $p(c|b)p(b|a)p(a)$. Setting $v_b$ as channel and marginalizing yields coarse-grained mechanism $\sum_{b}p(c|b)p(b|a)p(a) = p(c|a)p(a)$. The channel is rendered transparent and new mechanism $p(c|a)$ convolves $p(c|b)$ and $p(b|a)$.

\begin{step}
	Compute the effective graph of coarse-graining $X_\coarse$.
\end{step}

The micro-alphabet of unit $\mocc_l$ is $\tilde{\mAct}_l:=\prod_{k\in\mocc_l}\Act_k$.
The mechanism of $\mocc_l$ is computed as in Eq.~\eqref{e:u-mech} with the product restricted to occasions $j\in \chn\cup\mocc_l$, thus obtaining $p_{\mocc_l}(\act_l|\xin)$ where $\act_l\in\tilde{\mAct}_l$.

Two units $\mocc_k$ and $\mocc_l$ are connected by an edge if the outputs of $\mocc_k$ make a difference to the behavior of $\mocc_l$. More precisely, we draw an edge if $\exists \act_k,\act_k'\in \tilde{\mAct}_k$  such that
\begin{equation*}	
	p_{\mocc_l}(\act_l|\overline{\xin},\act_k)\neq p_{\mocc_l}(\act_l|\overline{\xin},\act_k')
	\mbox{ for some }\act_l\in \tilde{\mAct}_l. 
\end{equation*}
Here, $\overline{\xin}$ denotes the input from all units other than $\mocc_k$. 

The effective graph need not be acyclic. Intervening via the $do(-)$ calculus allows us to work with cycles.

\begin{step}
	Compute macro-alphabets of units in $X_\coarse$.
\end{step}

Coarse-graining can eliminate low-level details. Outputs that are distinguishable at the base level may not be after coarse-graining. This can occur in two ways. Outputs $b$ and $b'$ have indistinguishable effects if $p(a|b,c)=p(a|b',c)$ for all $a$ and $c$. Alternatively, two outputs react indistinguishably if $p(b|c)=p(b'|c)$ for all $c$.

More precisely, two outputs $u_l$ and $u_l'$ of unit $\mocc_l$ are equivalent, denoted $u_l\sim_{\coarse} u_l'$, iff
\begin{gather*}
	p_\coarse(\xout|\overline{\xin},u_l)=p_\coarse(\xout|\overline{\xin},u_l')\mbox{ and}\\
	p_{\mocc_l}(u_l|x_{in}^\coarse)=p_{\mocc_l}(u_l'|x_{in}^\coarse)\mbox{ for all }\xout, \xin.
\end{gather*}

Picking a single element from each equivalence class obtains the macro-alphabet $\mAct_l$ of the unit $\mocc_l$. The mechanism of $\mocc_l$ is $p_{\mocc_l}$, Step 4, restricted to macro-alphabets.

\section{Information}
\label{s:theory}

This section extends prior work to quantify the information generated by a cellular automaton, both as a whole and relative to its subsystems \citep{bt:08, bt:09}. 

Given subsystem $\fm$ of $\X$, let $p_{\fm}(\xout|\xin)$, or $\fm$ for short, denote its \emph{mechanism} or Markov matrix. The mechanism is computed by taking the Markov matrix of each occasion in $\X$, marginalizing over extrinsic inputs (edges not in $\X$) as in Eq.~\eqref{e:noise}, and taking the product. It is notationally convenient to write $p_\fm$ as though its inputs and outputs are $\xout$ and $\xin$, even though $\fm$ does not in general contain all occasions in $\X$ and therefore treats some inputs and outputs as extrinsic, unexplainable noise. We switch freely between terms ``subsystem'' and ``submechanism'' below.

\begin{figure}[thpb]
	\centering
	\includegraphics[scale=0.8]{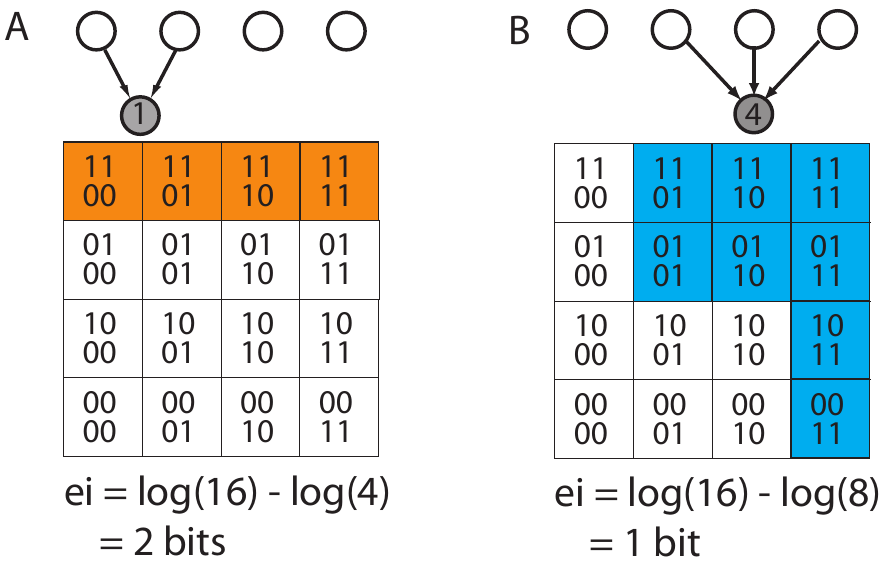}
	\caption{\footnotesize{
	Categorization and information.
	Cells fire if they receive two or more spikes. The $16=2^4$ possible outputs by the top layer are arranged in a grid.
	(AB): Cells $n_1$  and $n_4$ fire when the output is in the orange and blue regions respectively. Cell $n_1$'s response is more informative than $n_4$'s since it fires for fewer inputs.
	}}
	\label{f:information}
\end{figure}

\subsubsection{Effective information}
quantifies how selectively a mechanism discriminates between inputs when assigning them to an output. Alternatively, it measures how sharp the functional dependencies leading to an output are.

The \emph{actual repertoire} $\hat{p}_\fm(\Xin| \xout)$ is the set of inputs that cause (lead to) mechanism $\fm$ choosing output $\xout$, weighted by likelihood according to Bayes' rule
\begin{equation}
	\hat{p}_\fm\big(\xin|\xout\big) := \frac{p_\fm\big(\xout|do(\xin)\big)}{p(\xout)}\cdot p_{unif}(\xin).
	\label{e:arep}
\end{equation}

The $do(-)$ notation and hat $\hat{p}$ remind that we first \emph{intervene} to impose $\xin$ and then apply Markov matrix $p_\fm$. 

For deterministic mechanisms, i.e. functions $f:\Xin\rightarrow \Xout$,  the actual repertoire assigns $\hat{p}=\frac{1}{|f^{-1}(\xout)|}$ to elements of the pre-image and $\hat{p}=0$ to other elements of $\Xin$. The shaded regions in Fig.~\ref{f:information} show outputs of the top layer that cause the bottom cell to fire.

\emph{Effective information} generated when $\fm$ outputs $\xout$ is Kullback-Leibler divergence ($KL[p\|q]=\sum_i p_i\log_2\frac{p_i}{q_i}$),
\begin{equation}
	ei(\fm,\xout):= KL\Big[\hat{p}_\fm\big(\Xin|\xout\big)\Big\|p_{unif}(\Xin)\Big].
	\label{e:ei}
\end{equation}
Effective information is \emph{not} a statistical measure: it depends on the mechanism and a \emph{particular} output $\xout$.

Effective information generated by deterministic function $f$ is $ei(f,\xout)= \log_2 \frac{|\Xin|}{|f^{-1}(\xout)|}$ where $|\cdot|$ denotes cardinality. In Fig.~\ref{f:information}, $ei$ is the logarithm of the ratio of the total number of squares to the number of shaded squares.

\subsubsection{Excess information}
quantifies how much more information a mechanism generates than the sum of its submechanisms -- how synergistic the internal dependencies are. 

Given subsystem with mechanism $\fm$, partition $\cP=\{M^1\ldots M^m\}$ of the occasions in $src(\fm)$, and output $\xout$, define excess information as follows. Let $\fm^j:=\fm\cap (M^j\times X)$ be the restriction of $\fm$ to sources in $M^j$. \emph{Excess information} over $\cP$ is
\begin{equation}
	\xi(\fm,\cP,\xout) := ei(\fm,\xout) - \sum_{j}ei(\fm^j,\xout).
	\label{e:xip}
\end{equation}
Excess information (sans partition) is computed over the information-theoretic weakest link $\cP^{MIP}$
\begin{equation}
	\xi(\fm,\xout) := \xi(\fm,\cP^{MIP},\xout).
	\label{e:xi}
\end{equation}
Let $\Act_{M^j}:=\prod_{l\in M^j}\Act_j$. The minimum information partition\footnote{We restrict to bipartitions to reduce the computational burden.} $\cP^{MIP}$ minimizes normalized excess information:
\begin{gather*}
	\cP^{MIP}:=\arg\min_{\cP} \frac{\xi(\fm,\cP,\xout)}{\cN_\cP}, \mbox{ where}
	\\
	\cN_\cP:=(m-1)\cdot\min_j \left\{\log_2|\Act_{M^j}|\right\}.
\end{gather*}

Excess information is negative if any decomposition of the system generates more information than the whole.

\begin{figure}[thpb]
	\centering
	\includegraphics[scale=0.8]{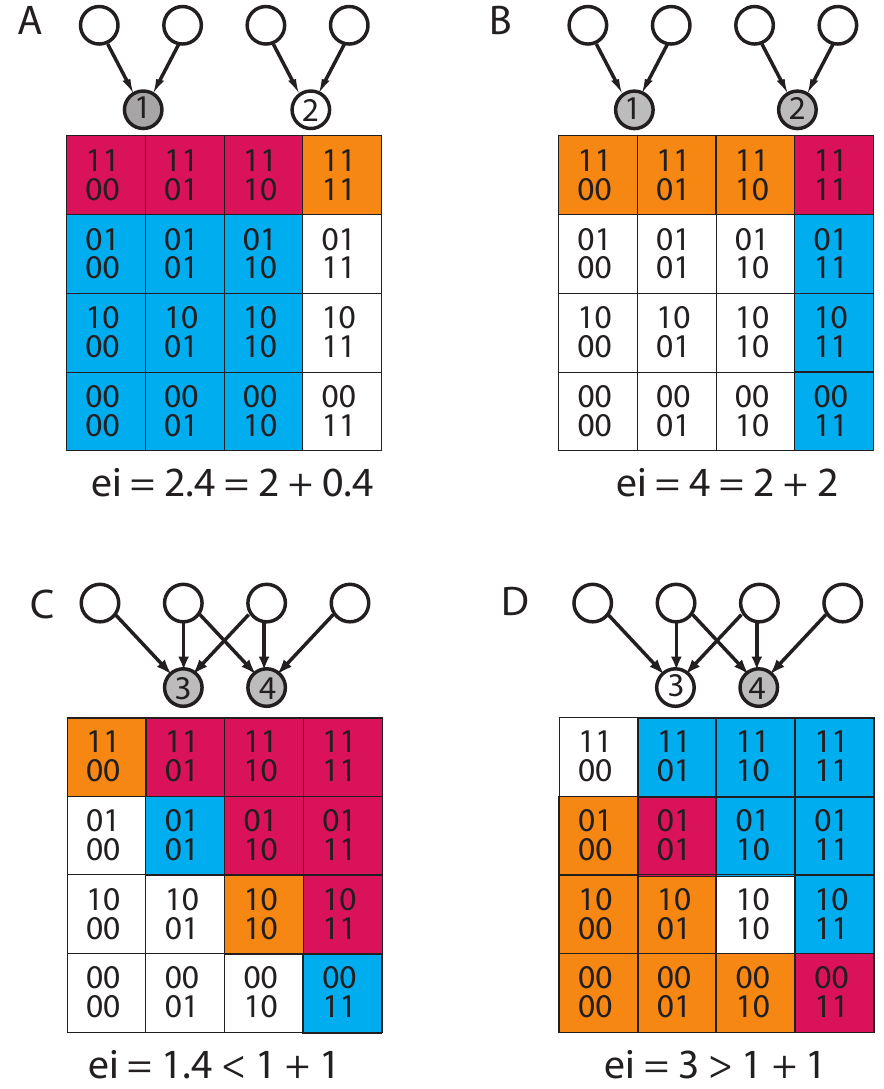}
	\caption{\footnotesize{
	Independent, redundant and synergistic information.
	(AB): Independent. Orthogonal categorizations, orange+pink and blue+pink shadings respectively, by $n_1$ and $n_2$. 
	(C): Partially redundant.  Both cells fire; categorizations overlap (pink) more ``than expected'' and $ei(n_3n_4,11)<ei(n_3,1)+ei(n_4,1)$. 
	(D): Synergistic. Overlap is less ``than expected''; $ei(n_3 n_4,01)>ei(n_3,0)+ei(n_4,1)$. 
	}}
	\label{f:integration}
\end{figure}

Fig.~\ref{f:integration} shows how two cells taken together can generate the same, less, or more information than their sum taken individually depending on how their categorizations overlap. Note the figure decomposes the mechanism of the system over \emph{targets} rather than sources and so does not depict excess information -- which is more useful but harder to illustrate.

Effective information and excess information can be computed for any submechanism of any coarse-graining of any cellular automaton. 

\begin{figure}[thpb]
	\centering
	\includegraphics[scale=0.8]{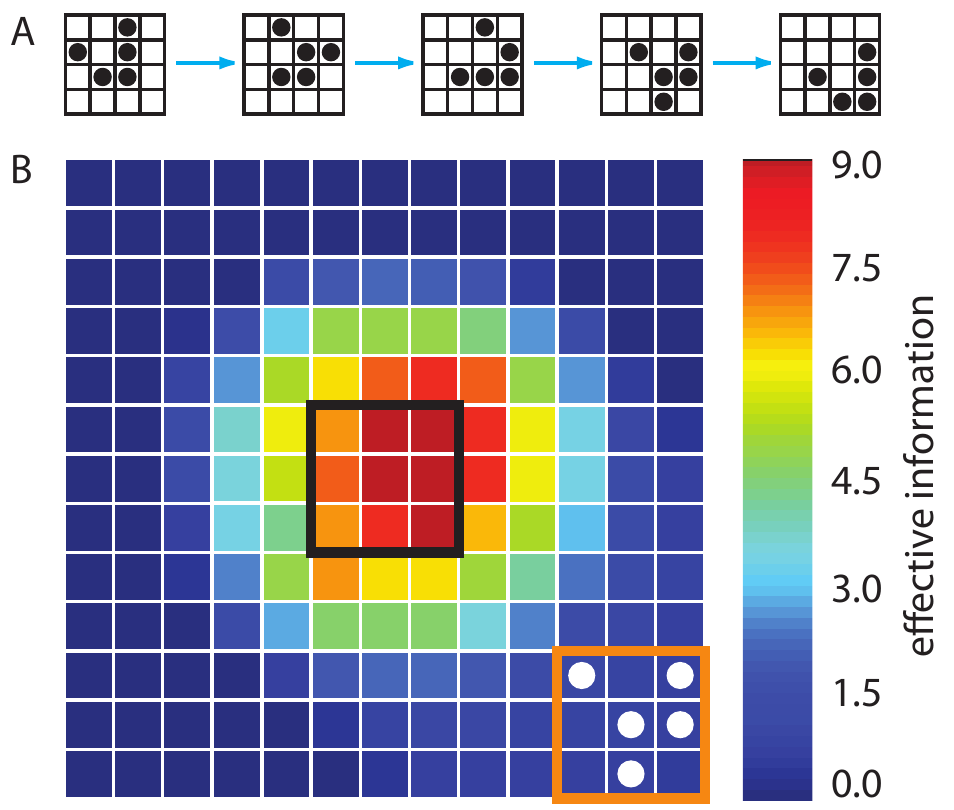}
	\caption{\footnotesize{
	Detecting focal points.
	(A): A glider moves 1 diagonal square every 4 time steps. 
	(B): Cells in the orange and black outlined $3\times3$ squares are units at $t=0$ and $t=-20$ respectively, with $\xout$ the glider shown. Cells at $t=-21$ are blank ground; other occasions are channel. Shifting the position of the black square produces a family of coarse-grainings. Effective information is shown as the black square's center varies over the grid.
	}}
	\label{f:framing}
\end{figure}

\section{Application: Conway's Game of Life}
\label{s:life}

The Game of Life has interesting dynamics at a range of spatiotemporal scales. At the atomic level, each coordinate (cell $i$ at time $t$) is an occasion and information processing is extremely local. At coarser granularities, information can propagate through channels, so that units generate information at a distance. Gliders, for example, are distributed objects that can interact over large distances in space and time, Fig.~\ref{f:framing}A, and provide an important example of an emergent process \citep{dennett:91, Beer:2004kl}. 

This section shows how effective and excess information quantifiably distinguish coarse-grainings expressing glider dynamics well from those expressing it badly.

\subsubsection{Effective information detects focal points.}
Fig.~\ref{f:framing}A shows a glider trajectory, which passes through 1 diagonal step over 4 tics. Fig.~\ref{f:framing}B investigates how glider trajectories are captured by coarse-grainings: if there is a glider in the $3\times3$ orange square at time 0, Fig.~\ref{f:framing}B, it must have passed through the black square at $t=-20$ to get there. Are coarse-grainings that respect glider trajectories quantifiably better than those that do not?

Fig.~\ref{f:framing}B fixes occasions in the black square at $t=-20$ and the orange square at $t=0$ as units (18 total), the ground as blank grid at $t=-21$ and everything else as channel. Varying the spatial location of the black square over the grid, we obtain a \emph{family} of coarse-grainings. Effective information for each graining in the family is shown in the figure. There is a clear focal point exactly where the black square intersects the spatiotemporal trajectory of the glider where $ei$ is maximized (dark red). Effective information is zero for locations that are too far or too close at $t=-20$ to effect the output of the orange square at $t=0$. 

Effective information thus provides a tool analogous to a camera focus: grainings closer to the focal point express glider dynamics better.

\begin{figure}[thpb]
	\centering
	\includegraphics[scale=0.8]{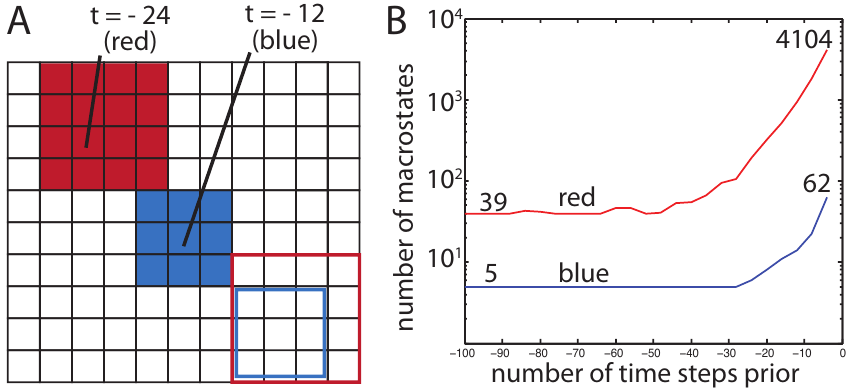}
	\caption{\footnotesize{
	Macro-alphabets as a function of distance.
	(A): Consider two families of coarse-grainings with channel and ground as in Fig.~\ref{f:framing}. First, take the blue squares (filled and empty) as units at times $-4n$ and $0$ where $n$ is the diagonal distance between them. Second, repeat for the red squares.
	(B): Log-plot of the size of the filled squares' macro-alphabets as a function of $-4n$.
	}}
	\label{f:texture}
\end{figure}

\subsubsection{Macroscopic texture varies with distance.}
The behavior of individual cells within a glider trajectory is far more complicated than the glider itself, which transitions through 4 phases as it traverses its diagonal trajectory, Fig.~\ref{f:framing}A. Does coarse-graining quantifiably simplify dynamics?

Fig.~\ref{f:texture} constructs pairs of $3\times3$ units out of occasions at various distances from one another and computes their macro-alphabets. A $3\times 3$ unit has a micro-alphabet of $2^9=512$ outputs. The macro-alphabet is found by grouping micro-outputs together into equivalences classes if their effect is the same after propagating through the channel. We find that the size of the macro-alphabet decreases exponentially as the distance between units increases, stabilizing at $5$ macro-outputs: the 4 glider phases in Fig.~\ref{f:framing}A and a large equivalence class of outputs that do not propagate to the target unit and are equivalent to a blank patch of grid. A similar phenomenon occurs for pairs of $4\times 4$ units, also Fig.~\ref{f:texture}.

Continuing the camera analogy: at close range the texture of units is visible. As the distance increases, the channel absorbs more of the detail. The computational texture of the system is simpler at coarser-grains yielding a more symbolic description where glider dynamics are described via 4 basic phases produced by a single macroscopic unit rather than $2^9$ outputs produced by 9 microscopic occasions. 

\begin{figure}[thpb]
	\centering
	\includegraphics[scale=0.8]{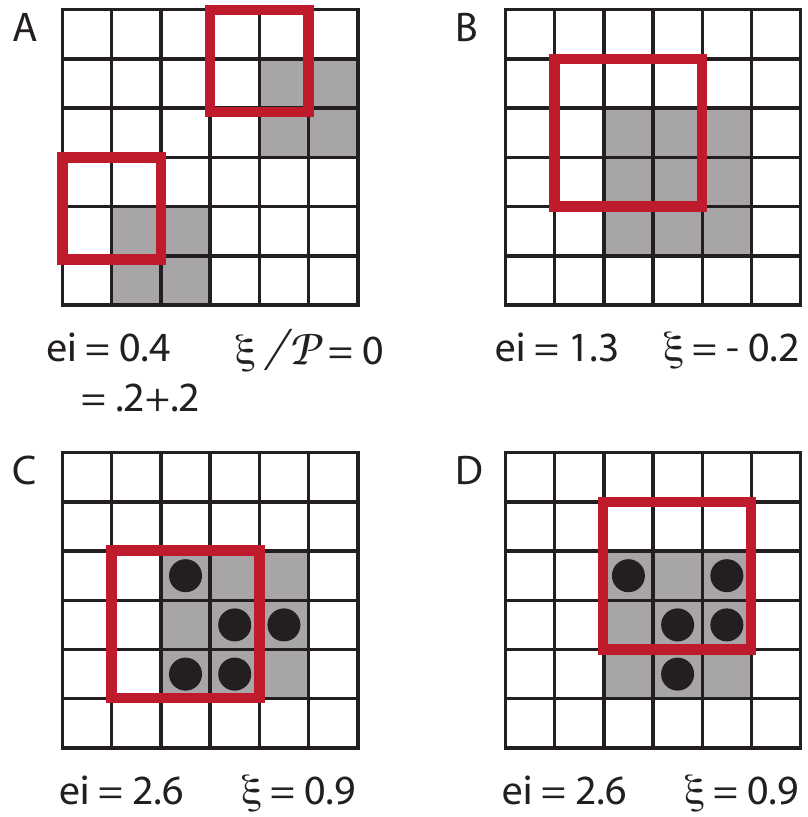}
	\caption{\footnotesize{
	Detecting spatial organization. Units are the cells in the red (thick-edged) and gray (filled) squares at $t=0$ and $t=1$ respectively; other occasions are extrinsic noise.
	(A): $\xi=0$. The coarse-graining groups non-interacting occasions into units. 
	(B): $\xi<0$. A blank grid is highly redundant.
	(CD): $\xi>0$. Gliders perform interesting information-processing.
	}}
	\label{f:chunking}
\end{figure}

\subsubsection{Excess information detects spatial organization.}
So far we have only considered grainings of the Game of Life that respect its spatial organization -- in effect, taking the spatial structure for granted. \emph{A priori}, there is nothing stopping us from grouping the 8 gray cells in Fig.~\ref{f:chunking}A into a single unit that \emph{does not} respect the spatial organization, since its constituents are separated in space. Are coarse-grainings that respect the grid-structure quantifiably better than others?

Fig.~\ref{f:chunking}A shows a coarse-graining that does \emph{not} respect the grid. It constructs two units, one from \emph{both} gray squares at $t=1$ and the other from \emph{both} red squares at $t=0$. Intuitively, the coarse-graining is unsatisfactory since it builds units whose constituent occasions have nothing to do with each other over the time-scale in question. Quantitatively, excess information over the obvious partition $\cP$ of the system into two parts is 0 bits. It is easy to show $\xi\leq 0$ for any disjoint units. By comparison, the coarse-grainings in panels CD, which respect the grid structure, both generate positive excess information. 

Thus we find that not only does our information-theoretic camera have an automatic focus, it also detects when processes hang together to form a single coherent scene.

\subsubsection{Excess information detects gliders.}
Blank stretches of grid, Fig.~\ref{f:chunking}B, are boring. There is nothing going on. Are interesting patches of grid quantifiably distinguishable from boring patches?

Excess information distinguishes blank grids from gliders: $\xi$ on the blank grid is negative, Fig.~\ref{f:chunking}B , since the information generated by the cells is redundant analogous to Fig.~\ref{f:integration}C. By contrast, $\xi$ for a glider is positive, Fig.~\ref{f:chunking}CD, since its cells perform synergistic categorizations, similarly to Fig.~\ref{f:integration}D. Glider trajectories are also captured by excess information: varying the location of the red units (at $t=0$) around the gray units we find that  $\xi$ is maximized in the positions shown, Fig.~\ref{f:chunking}CD, thus expressing the rightwards and downwards motions of the respective gliders.

Returning to the camera analogy, blank patches of grid fade into (back)ground or are (transparent) channel, whereas gliders are highlighted front and center as units.

\section{Application: Hopfield networks}
\label{s:hopfield}


Hopfield networks embed energy landscapes into their connectivity. For any initial condition they tend to one of few attractors -- troughs in the landscape  \citep{hopfield:82, amit:89}. Although cells in Hopfield networks are quite different from neurons, there is evidence suggesting neuronal populations transition between coherent distributed states similar to attractors \citep{abeles:95, Jones:2007xy}. 

Attractors are population level phenomena. They arise because of interactions between groups of cells -- no single cell is responsible for their existence -- suggesting that coarse-graining may reveal interesting features of attractor dynamics. 

\subsubsection{Effective information detects causal interactions.}
Table~\ref{t:transitions} analyzes a sample run of unidirectionally coupled Hopfield networks $A\rightarrow B$. Network $A$ is initialized at an unstable point in the energy landscape and $B$ in an attractor. $A$ settles into a different attractor from $B$ and then shoves $B$ into the new attractor over a few time steps. Intuitively, $A$ only exerts a strong force on $B$ once it has settled in an attractor and before $B$ transitions to the same attractor. Is the force $A$ exerts on $B$ quantitatively detectable?

Table \ref{t:transitions} shows the effects of $A$ and $B$ respectively on $B$ by computing $ei$ for two coarse-grainings constructed for each transition $t\rightarrow t+1$. Coarse-graining INT sets cells in  $B$ at $t$ and $t+1$ as units and $A$ as extrinsic noise. EXT sets cells in $A$ at $t$ and $B$ at $t+1$ as units and fixes $B$ at time $t$ as ground.

INT generates higher $ei$ for all transitions except $1\rightarrow2\rightarrow3$, precisely  when $A$ shoves $B$. Effective information is high when an output is sensitive to changes in an input so it is unsurprising that $B$ is more sensitive to changes in $A$ exactly when $A$ forces $B$ out from one attractor into another. Analyzing other sample runs (not shown) confirms that $ei$ reliably detects when $A$ shoves $B$ out of an attractor.

\begin{table}[t]
    \begin{center}
    \begin{small}
    \begin{tabular}{@{\extracolsep{\fill}} c | c c | c c | c c }
	& \multicolumn{2}{c}{output} 
	& \multicolumn{2}{c}{\hspace{-1mm}INT: $B\rightarrow B$}
	& \multicolumn{2}{c}{\hspace{-1mm}EXT: $A\rightarrow B$}
	\vspace{-4mm}\\ 
	\toprule   & & & & & \\
	$t$ & $A$        & $B$        & $ei$            & $\max\xi$       & $ei$            & $\max\xi$       \\
	 0  & $00000000$ & $01010101$ &                 &                 &                 &                 \\
	 1  & $10100011$ & $01010101$ & $\mathbf{2.42}$ & $\mathbf{0.10}$ & $0.31$          & $0.04$          \\
	 2  & $10101010$ & $00010101$ & $1.85$          & $0.08$          & $\mathbf{2.44}$ & $\mathbf{0.16}$ \\
	 3  & $10101010$ & $00101011$ & $1.96$          & $0.12$          & $\mathbf{6.89}$ & $\mathbf{0.27}$ \\
	 4  & $10101010$ & $00101010$ & $\mathbf{1.85}$ & $0.08$          & $1.60$          & $\mathbf{0.10}$ \\
	 5  & $10101010$ & $10101010$ & $\mathbf{2.42}$ & $\mathbf{0.10}$ & $0.90$          & $0.06$          \\
	 6  & $10101010$ & $10101010$ & $\mathbf{2.42}$ & $\mathbf{0.10}$ & $0.31$          & $0.04$          \\
	\bottomrule
    \end{tabular}
    \end{small}
    \end{center}
    \vskip -0.1in
    \caption{\footnotesize{
	Analysis of unidirectionally coupled Hopfield networks $A\rightarrow B$ each containing 8 cells. The networks and coupling embed attractors $\{00001111, 00110011, 01010101\}$ and their mirrors.  Temperature is $T=0.25$. A sample run is analyzed using two coarse-grainings: INT captures $B$'s effect on itself and EXT captures $A$'s effect on B; see text.
	}}
    \label{t:transitions}
    \vskip -0.1in
\end{table}

\subsubsection{Macroscopic mechanisms depend on the ground.}
Fixing the ground incorporates population-level biases into a coarse-grained cellular automaton's information-processing.

The ground in coarse-graining EXT (i.e. the output of $B$ at $t-1$) biases the mechanisms of the units in $B$ at time $t$. When the ground is an attractor, it introduces tremendous inertia into the coarse-grained dynamics since $B$ is heavily biased towards outputting the attractor again. Few inputs from $A$ can overcome this inertia, so if $B$ is pushed out of an attractor it generates high $ei$ about $A$. Conversely, when $B$ stays in an attractor, e.g. transition $5\rightarrow 6$, it follows its internal bias and so generates low $ei$ about $A$.

\subsubsection{Excess information detects attractor redundancy.}
Following our analysis of gliders, we investigate how attractors are captured by excess information. It turns out that $\xi$ is negative in all cases: the functional dependencies within Hopfield networks are redundant. An attractor is analogous to a blank Game of Life grid where little is going on. Thus, although attractors are population-level phenomena, we exclude them as emergent processes. 

\subsubsection{Excess information expresses attractor transitions.}
We therefore refine our analysis and compute the subset of units at time $t$ that maximize $\xi$; maximum values are shown in Table \ref{t:transitions}. We find that the system decomposes into pairs of occasions with low $\xi$, except when $B$ is shoved, in which case larger structures of 5 occasions emerge. This fits prior analysis showing transitions between attractors yield more integrated dynamics \citep{bt:08} and suggestions that cortical dynamics is metastable, characterized by antagonism between local attractors \citep{friston:97}.

Our analysis suggests that \emph{transitions} between attractors are the most interesting emergent behaviors in coupled Hopfield networks. How this generalizes to more sophisticated models remains to be seen.

\section{Emergence}
\label{s:emergence}

The examples show we can quantify how well a graining expresses a cellular automaton's dynamics. Effective information detects glider trajectories and also captures when one Hopfield network shoves another. However, $ei$ does not detect whether a unit is integrated. For this we need excess information, which compares the information generated by a mechanism to that generated by its submechanisms. Forming units out of disjoint collections of occasions yields $\xi=0$. Moreover, boring units (such as blank patches of grid or dead-end fixed point attractors) have negative $\xi$. Thus, $\xi$ is a promising candidate for quantifying emergent processes.

This section formalizes the intuition that a system is emergent if its dynamics are better expressed at coarser spatiotemporal granularities.  The idea is simple. Emergent units should generate more excess information, and have more excess information generated about them, than their sub-units. Moreover emergent units should generate more excess information than \emph{neighboring} units, recall Fig.~\ref{f:framing}.

Stating the definition precisely requires some notation. Let $\mathfrak{src}_{\occ_l}=\{\occ_l\}\cup\{\occ_k|k\rightarrow l\}$ and similarly for $\mathfrak{trg}_{\occ_l}$. Let $\mathcal {J}$ be a subgraining of $\coarse$, denoted ${\mathcal J}\prec \coarse$, if for every $\mocc_j\in{\mathcal J}$ there is a unit $\mocc_k\in\coarse$ such that $\mocc_j\subsetneq \mocc_k$. We compare mechanism $\fm\subset\coarse$ with its subgrains via
\begin{equation*}
	\xi_{\coarse/{\mathcal J}}(\fm,\xout):=ei_{\tilde{\coarse}}(\fm,\xout)
	- \sum_{\occ_j\in{\mathcal J}} ei_{\tilde{\mathcal J}}(\fm^j,\xout),
\end{equation*}
where $\fm^j=\fm\cap\mathfrak{src}_{\occ_j}$ and $ei_{\tilde{\coarse}}$ signifies effective information is computed over $\coarse$ using micro-alphabets.

\begin{defn}[emergence]
	Fix cellular automaton $\X$ with output $\xout$. Coarse-graining\footnote{Ground output $s^\grd$ is $\xout$ restricted to ground occasions.} $\coarse$ is \emph{emergent} if it satisfies conditions \emph{E1} and \emph{E2}.
\end{defn}
\begin{enumerate}[E1.]
	\item	
	Each unit $\mocc_l\in\coarse$ generates excess information about its sources and has excess information generated about it by its targets, relative to subgrains ${\mathcal J}\prec\coarse$:
\begin{equation}
	0 < \xi_{{\mathcal J}/\coarse}\big({\mathfrak{src}}_{\mocc_l},\xout\big)
	\mbox{ and } 
	0 < \xi_{{\mathcal J}/\coarse}\big({\mathfrak{trg}}_{\mocc_l}, \xout\big).
	\label{e:src_trg}
\end{equation}
	\item
	There is an \emph{emergent} subgrain ${\mathcal J}\prec\coarse$ such that (i) every unit of $\coarse$ contains a unit of ${\mathcal J}$ and (ii) neighbors $\coarse'$ (defined below) of $\coarse$ with respect to ${\mathcal J}$ satisfy 
	\begin{equation}
		\xi_{{\mathcal J}/\coarse'}\big({\mathfrak{src}}_{\mocc'},\xout\big)
		\leq \xi_{{\mathcal J}/\coarse}\big({\mathfrak{src}}_{\mocc},\xout\big)
		\label{e:nbrs}
	\end{equation}
	for all $\mocc\in\coarse$, and similarly for $\mathfrak{trg}$'s.
\end{enumerate}
If $\coarse$ has no emergent subgrains then E2 is vacuous.

Grain $\coarse'$ is a \emph{neighbor} of $\coarse$ with respect to ${\mathcal J}\prec \coarse$ if for every $\mocc\in\coarse$ there is a unique $\mocc'\in\coarse'$ satisfying
\begin{enumerate}[N1.]
	\item there is a unit $T\in{\mathcal J}$ such that
	$T\subset\mocc, \mocc'$, $\mathfrak{src}_T\subset \mathfrak{src}_{\mocc},\mathfrak{src}_{\mocc'}$ and similarly for $\mathfrak{trg}$; and
	\item the alphabet of $\mocc'$ is no larger than $\mocc$: $\left|\prod_{k\in\mocc'}\Act_k\right|\leq \left|\prod_{l\in\mocc}\Act_l\right|$, and similarly for the combined alphabets of their sources and targets respectively.
\end{enumerate}

The graining ${\mathcal E}_\X$ that best expresses  $\X$ outputting $\xout$ is found by maximizing normalized excess information:
\begin{equation}
	{\mathcal E}_X(\xout) := \arg\max_{\{\coarse\,|\,\mbox{emergent}\}} 
	\frac{\xi(\coarse,\xout)}{{\mathcal N}_{{\mathcal P}^{MIP}}^\coarse}.
	\label{e:max}
\end{equation}
Here, ${\mathcal N}_{{\mathcal P}^{MIP}}^\coarse$ is the normalizing constant found when computing the minimum information partition for  $\coarse$.

\subsubsection{Some implications.}
We apply the definition to the Game of Life to gain insight into its mechanics.

Condition E1 requires that interactions between units and their sources (and targets) are synergistic, Fig.~\ref{f:chunking}CD. Units that decompose into independent pieces, Fig.~\ref{f:chunking}A, or perform highly redundant operations, Fig.~\ref{f:chunking}B, are therefore not emergent. 

Condition E2 compares units to their neighbors. Rather than build the automaton's spatial organization directly into the definition, neighbors of $\coarse$ are defined as coarse-grainings whose units overlap with $\coarse$ and whose alphabets are no bigger. Coarse-grainings with higher $\xi$ than their neighbors are closer to focal points, recall Fig.~\ref{f:framing} and Fig.~\ref{f:chunking}CD, where $\xi$ was maximized for units respecting glider trajectories. An analysis of glider boundaries similar in spirit to this paper is \citep{Beer:2004kl}.

Finally, Eq.~\eqref{e:max} picks out the most expressive coarse-graining. The normalization plays two roles. First, it biases the optimization towards grainings whose MIPs contain few, symmetric parts following \citep{bt:08}. Second, it biases the optimization towards systems with simpler macro-alphabets. Recall, Fig.~\ref{f:texture}, that coarse-graining produces more symbolic interactions by decreasing the size of alphabets. Simplifying alphabets typically reduces effective and excess information since there are less bits to go around. The normalization term rewards simpler levels of description, so long as they use the bits in play more synergistically.

\section{Discussion}
\label{s:discussion}

In this paper we introduced a flexible, scalable coarse-graining method that applies to any cellular automaton. Our notion of automaton applies to a broad range of systems. The constraints are that they (i) decompose into discrete components with (ii) finite alphabets where (iii) time passes in discrete tics. We then described how to quantify the information generated when a system produces an output (at any scale) both as a whole and relative to its subsystems. An important feature of our approach is  that the output $\xout$ of a graining is incorporated into the ground and also directly influences $ei$ and $\xi$ through computation of the actual repertoires. Coarse-graining and emergence therefore capture some of the \emph{suppleness} of biological processes \citep{Bedau:1997tg}: they are context-dependent and require many \emph{ceteris paribus} clauses (i.e. background) to describe. 

Investigating examples taken from Conway's Game of Life and coupled Hopfield networks, we accumulated a small but significant body of evidence confirming the principle that \emph{expressive coarse-grainings generate more information relative to sub-grainings}. Finally, we provisionally defined emergent processes. The definition is provisional since it derives from analyzing a small fraction of the possible coarse-grainings of only two kinds of cellular automata.

Hopfield networks and the Game of Life are simple models capturing some important aspects of biological systems. Ultimately, we would like to analyze emergent phenomena in more realistic models, in particular of the brain. Conscious percepts take 100-200ms to arise and brain activity is (presumably) better expressed as comparatively leisurely interactions between neurons or neuronal assemblies rather than much faster interactions between atoms or molecules \citep{tononi:04}. To apply the techniques developed here to more realistic models we must confront a computational hurdle: the number of coarse-grainings that can be imposed on large cellular automata is vast. Nevertheless, the approach developed here may still be of use. First, manipulating macro-alphabets provides a method for performing approximate computations on large-scale systems. Second, for more fine-grained analysis, initial estimates about which coarse-grainings best express a system's dynamics can be fine-tuned by comparing them with neighbors. 

\subsubsection{Acknowledgements.}
The author thanks Dominik Janzing for many useful comments on an earlier draft, Giulio Tononi for stimulating conversations and Virgil Griffiths for emphasizing the importance of excess information.

{
\footnotesize{

}


\begin{thebibliography}{}

\bibitem[Abeles et~al., 1995]{abeles:95}
Abeles, M., Bergman, H., Gat, I., Meilijson, I., Seidemann, E., Tishby, N., and
  Vaadia, E. (1995).
\newblock Cortical activity flips among quasi-stationary states.
\newblock {\em Proc. Nat. Acad. Sci.}, 92:8616--8620.

\bibitem[Amit, 1989]{amit:89}
Amit, D. (1989).
\newblock {\em Modelling brain function: the world of attractor neural
  networks}.
\newblock Cambridge University Press.

\bibitem[Anderson, 1972]{Anderson:1972oq}
Anderson, P.~W. (1972).
\newblock More is different.
\newblock {\em Science}, 177(4047):393--6.

\bibitem[Balduzzi and Tononi, 2008]{bt:08}
Balduzzi, D. and Tononi, G. (2008).
\newblock Integrated {I}nformation in {D}iscrete {D}ynamical {S}ystems:
  {M}otivation and {T}heoretical {F}ramework.
\newblock {\em PLoS Comput Biol}, 4(6):e1000091.

\bibitem[Balduzzi and Tononi, 2009]{bt:09}
Balduzzi, D. and Tononi, G. (2009).
\newblock Qualia: the geometry of integrated information.
\newblock {\em PLoS Comput Biol}, 5(8):e1000462.

\bibitem[Bedau, 1997]{Bedau:1997tg}
Bedau, M.~A. (1997).
\newblock Emergent models of supple dynamics in life and mind.
\newblock {\em Brain Cogn}, 34(1):5--27.

\bibitem[Beer, 2004]{Beer:2004kl}
Beer, R.~D. (2004).
\newblock Autopoiesis and cognition in the game of life.
\newblock {\em Artif Life}, 10(3):309--26.

\bibitem[Berlekamp et~al., 1982]{berlekamp:82}
Berlekamp, E., Conway, J., and Guy, R. (1982).
\newblock {\em Winning Ways for your Mathematical Plays}, volume~2.
\newblock Academic Press.

\bibitem[Crutchfield, 1994]{crutchfield:94}
Crutchfield, J. (1994).
\newblock The calculi of emergence: {C}omputation, dynamics, and induction.
\newblock {\em Physica D}, 75:11--54.

\bibitem[Dennett, 1991]{dennett:91}
Dennett, D.~C. (1991).
\newblock Real {P}atterns.
\newblock {\em J. Philosophy}, 88(1):27--51.

\bibitem[Friston, 1997]{friston:97}
Friston, K. (1997).
\newblock Transients, metastability and neuronal dynamics.
\newblock {\em Neuroimage}, 5:164--171.

\bibitem[Hopfield, 1982]{hopfield:82}
Hopfield, J. (1982).
\newblock Neural networks and physical systems with emergent computational
  properties.
\newblock {\em Proc. Nat. Acad. Sci.}, 79:2554--2558.

\bibitem[Jones et~al., 2007]{Jones:2007xy}
Jones, L.~M., Fontanini, A., Sadacca, B.~F., Miller, P., and Katz, D.~B.
  (2007).
\newblock Natural stimuli evoke dynamic sequences of states in sensory cortical
  ensembles.
\newblock {\em Proc Natl Acad Sci U S A}, 104(47):18772--18777.

\bibitem[Pearl, 2000]{pearl:00}
Pearl, J. (2000).
\newblock {\em Causality: models, reasoning and inference}.
\newblock Cambridge University Press.

\bibitem[Polani, 2006]{polani:06}
Polani, D. (2006).
\newblock Emergence, intrinsic structure of information, and agenthood.
\newblock {\em Int J Complex Systems}, 1937.

\bibitem[Seth, 2010]{Seth:2010ve}
Seth, A.~K. (2010).
\newblock Measuring autonomy and emergence via {G}ranger causality.
\newblock {\em Artif Life}, 16(2):179--96.

\bibitem[Shalizi and Moore, 2006]{shalizi:06}
Shalizi, C. and Moore, C. (2006).
\newblock What is a macrostate: {S}ubjective observations and objective
  dynamics.
\newblock {\em http://arxiv.org/abs/condmat/0303625}.

\bibitem[Tononi, 2004]{tononi:04}
Tononi, G. (2004).
\newblock An information integration theory of consciousness.
\newblock {\em BMC Neurosci}, 5:42.

\end{thebibliography}
\end{document}